\documentclass{article}
\usepackage{url}
\usepackage{multicol}
\usepackage{amsmath}
\usepackage[english]{babel}
\usepackage{graphicx}
\usepackage{subfigure}
\usepackage{tikz}
\usepackage{lipsum}
\usepackage{caption}
\usepackage{algorithm}
\usepackage{algorithmic}
\usepackage{listings}
\usepackage{ifthen}
\usepackage{booktabs, caption, fixltx2e}
\usepackage[flushleft]{threeparttable}
\newcounter{algorithmus}[section]

\newcounter{algzeilennummer}[algorithmus]

\newlength{\algeinrueckung}
\newlength{\algbreite}
\newlength{\boxbreite}
\newlength{\algeinrueckungdelta}
\newboolean{alginzeile}

\newcommand{\algln}[1][]{\refstepcounter{algzeilennummer}%
  \ifthenelse{\equal{#1}{}}{}{\label{#1}}%
  \ifthenelse{\boolean{alginzeile}}{\end{flushleft}\end{minipage}\newline}{}%
  \makebox[\boxbreite][r]{(\arabic{algzeilennummer})}%
  \hspace{\algeinrueckung}
  \begin{minipage}[t]{\algbreite}\begin{flushleft}%
  \setboolean{alginzeile}{true}
}

\newcommand{\rein}{
  \ifthenelse{\boolean{alginzeile}}{\end{flushleft}\end{minipage}\newline\setboolean{alginzeile}{false}}{}%
  \addtolength{\algeinrueckung}{\algeinrueckungdelta}
  \addtolength{\algbreite}{-\algeinrueckungdelta}
}

\newcommand{\raus}{
  \ifthenelse{\boolean{alginzeile}}{\end{flushleft}\end{minipage}\newline\setboolean{alginzeile}{false}}{}%
  \addtolength{\algeinrueckung}{-\algeinrueckungdelta}
  \addtolength{\algbreite}{\algeinrueckungdelta}
}
\makeatletter
\newcommand\figcaption{\def\@captype{figure}\caption}
\makeatother
\newtheorem{example}{Example}%[section]
\newtheorem{definition}{Definition}%[section]
\title{Synthesis of Linear Nearest Neighbor Quantum Circuits}
\author{Md.\ Mazder Rahman \& Gerhard W.\ Dueck\\University of New Brunswick\\Canada\\Mazder.Rahman@unb.ca \& gdueck@unb.ca}
\date{}
\begin{document}
\maketitle
\thispagestyle{empty}

Presented at the 10th International Workshop on Boolean Problems (2012), Freiberg, Germany.

\begin{abstract}
%The text of the abstract is indented 10mm to the left and right. 
%The font is Times New Roman 9pt. The abstract should not be longer than 100 words.
This paper presents  models for transforming standard reversible circuits into Linear Nearest Neighbor (LNN) architecture without inserting SWAP gates. Templates to optimize the transformed LNN circuits are proposed. All minimal LNN circuits for all 3-qubit functions have been generated to serve as benchmarks to evaluate heuristic optimization algorithms. The minimal results generated are compared with optimized LNN circuits obtained from the post synthesis algorithm --- template matching with LNN templates. Experiments show that the suggested synthesis flow significantly improves the quantum cost of circuits. 
\end{abstract}

\begin{section}{Introduction}\label{sec:intro}
%The inherent property of reversibility as well as the infinite state space in quantum computation~\cite{MC:2000} leads to synthesize reversible logic into quantum circuits.
For the last decades, significant research on synthesizing quantum circuits has been done. %These synthesis approaches are based on gates that are realizable on any qubits even if these qubits are located far away in physical space~\cite{Mper:2011}. 
Most synthesis approaches ignore physical constrains, i.e. operation may be applied to qubits that are distant in physical space~\cite{Mper:2011}. 
However, some technologies such as one dimensional Ion Trap only support the Linear Nearest Neighbor (LNN) architecture of circuits in which the control and target of a gate must be adjacent. Therefore, the synthesis of LNN circuits is of interest. % in active research.  
%The direct synthesis of both standard quantum circuits and LNN circuits are intractable, therefore, synthesis heuristics such as~\cite{MMD:2003}, \cite{IKY:2002}, \cite{MP:2002} based on Toffoli are used to obtain reversible circuits from the binary reversible function specifications. 
Toffoli networks can be transformed into LNN quantum circuits by using the standard decomposition of multiple-control Toffoli (MCT) circuits~\cite{BBC+:1995} and further inserting SWAP gates~\cite{Robert:2011} or appropriate SWAP sequence~\cite{ZMismvl+:2011} whenever a gate with non-adjacent control and target occurs. The obtained  circuits are optimized by post synthesis methods. One such method is template matching with \textbf{SWAP templates} proposed in~\cite{Robert:2011}. 
%However, this results in LNN circuits with considerably high quantum cost even in the case of smaller circuits. 
 In this paper, we identify efficient ways of transforming standard MCT circuits into LNN architecture. 
\end{section}

\begin{section}{Background}\label{sec:liter_review}
%\section{Preliminaries} \label{Preliminaries}
%To keep the paper self-contained, background of reversible circuits, logic operations in quantum circuits as well as quantum templates are overviewed in this section. 
%%%%%%%%%%%%%%%%%%%%%%%%%%%%%%%%%%%%%%%%
% it was not in previous version
A Boolean logic function $f:B^n\rightarrow B^n$ is said to be reversible if there is a one-to-one and onto mapping between input vectors and output vectors. A reversible function can be embedded into a \textbf{reversible circuit} by cascading the \textbf{reversible gates}
% such as  generalized Multiple Control Toffoli $(MCT)$,  Peres~\cite{PhysRevA.32.3266}  and Fredkin\cite{ar:ft} 
without allowing feedback and fanout to preserve the reversibility. %However, the synthesis of reversible logic based on generalized Multiple Control Toffoli $(MCT)$ is prevalent and the resulting circuits are referred to as $MCT$ circuits. However, the
A generalized  multiple-control Toffoli gate is defined as $T_n(C,t)$ based on number of  lines $0 < n$,  which maps the pattern $(x_{i_1},x_{i_2},...,x_{i_k})$ to $(x_{i_1},x_{i_2},...,x_{j-1},x_j\oplus x_{i_1}x_{i_2} \ldots x_{j-1} x_{j+1} \ldots x_{i_k} , x_{j+1}, \ldots , x_{i_k})$, where $C=\{x_{i_1},x_{i_2},...,x_{i_k}\}$, $t=\{x_j\}$ and $C \cap t=\phi$. $C$ is referred to as the control set and $t$ is referred to as the target. $T_1$ and $T_2$ are referred to as $NOT$ and $CNOT$ respectively. A picture of a $T_3$ gate is shown in Figure~\ref{fig:k} \subref{fig:t32}.

The Controlled-$V$ gate has two lines (control and target), the target
line changes using the transformation defined by the matrix $V = \frac{i+1}{2}  \bigr( \begin{smallmatrix} 1 & -i \\ -i & 1 \end{smallmatrix} \bigr) $ if the control line has the value 1.
similarly, the Controlled-$V^\dagger$ gate has two lines (control and target), the target
line changes using the transformation defined by the matrix $V^\dagger = V^{-1} = \frac{i-1}{2} \bigr( \begin{smallmatrix} 1 & i \\ i & 1 \end{smallmatrix} \bigr) $ if the control line has the value 1.
The SWAP$(x,y)$ gate maps the input $(x,y)$ to $(y,x)$.

Logic operations in quantum computation are quite different from those in classical logic. The fundamental unit of information in quantum computation is a qubit represented by a state vector. A qubit has a state either $|0\rangle$ or $|1\rangle$ these are known as computational basis states.  An arbitrary qubit is described by the following state  vector 
\begin{equation}
   |\psi\rangle=\alpha|0\rangle+\beta|1\rangle=\left(\begin{array}{c}
  \alpha\\
  \beta\end{array}\right)
\end{equation}
where $\alpha$ and $\beta$ are complex numbers that satisfy the constraint $|\alpha|^{2}+|\beta|^{2}=1$. 
The measurement of a qubit results either 0 with probability $|\alpha|^2$, that is, the state  $|0\rangle=\bigl(\begin{smallmatrix}
1\\ 0
\end{smallmatrix} \bigr)$    
or $1$ with probability $|\beta|^2$, that is, the state $|1\rangle=\bigl(\begin{smallmatrix}
0\\ 1
\end{smallmatrix} \bigr)$. On the other hand, a classical bit has a state either $0$ or $1$ which is analogous to the measurement of a qubit state either  $|0\rangle$ or $|1\rangle$ respectively. The fundamental difference between bits and qubits is that a bit can be either state $0$ or $1$ whereas a qubit can be a state rather than $|0\rangle$ or $|1\rangle$. A two qubit system has four computation basis states $|00\rangle$, $|01\rangle$, $|10\rangle$ and $|11\rangle$ can be represented by the sate vector

\begin{equation}|\psi\rangle= \lambda_{1}|00\rangle+\lambda_{2}|01\rangle+\lambda_{3}|10\rangle+\lambda_{4}|11\rangle=\left(\begin{array}{c}
\lambda_{1}\\
\lambda_{2}\\
\lambda_{3}\\
\lambda_{4}\end{array}\right)\end{equation} where$\lambda_{1}\lambda_{4} = \lambda_{2}\lambda_{3}$. If $\lambda_{1}\lambda_{4}\neq \lambda_{2}\lambda_{3}$ then the state $|\psi\rangle$ is referred to as an \textbf{entangled state} which is not separable as the tensor product of two single qubits.
% In addition, it is possible to form linear combination of states$-$so called superpositions, therefore, a two-qubit state can be formed by the tensor product of two single qubit states $\alpha_{1}|0\rangle+\beta_{1}|1\rangle$ and $\alpha_{2}|0\rangle+\beta_{2}|1\rangle$  represented  as \[
%\left(\begin{array}{c}
%\alpha_{1}\\
%\beta_{1}\end{array}\right)\otimes\left(\begin{array}{c}
%\alpha_{2}\\
%\beta_{2}\end{array}\right)=\left(\begin{array}{c}
%\alpha_{1}\alpha_{2}\\
%\alpha_{1}\beta_{2}\\
%\beta_{1}\alpha_{2}\\
%\beta_{1}\beta_{2}\end{array}\right)\] where $\alpha_{1}\alpha_{2}\beta_{1}\beta_{2}=\alpha_{1}\beta_{2}\beta_{1}\alpha_{2}$, otherwise the resulting state is entangled.
%%%%%%%%%%%%%%%%%%%%%%%%%%%%%%%%%%%%%%%%%%%%%%%%
The elementary quantum gates $NOT$, $CNOT$, Controlled-$V$ and Controlled-$V^\dagger$ are also known as quantum primitives have been widely used to synthesis of binary reversible functions. 
%(for a formal definition of these gates, please see \cite{BBC+:1995}).  
A \textbf{quantum circuit} is realized by the cascades of quantum primitives. The \textbf{quantum cost} of a reversible circuit is defined by the number of quantum gates required to realized the circuit. To perform the logic operations in quantum circuits, two more qubit states  $|v_0\rangle$ and $|v_1\rangle$  rather than $|0\rangle$, $|1\rangle$, are possible at the intermediate position in the circuits where  
$|v_0\rangle=\frac{(1+i)}{2}\bigl(\begin{smallmatrix}
1\\ -i
\end{smallmatrix} \bigr)$ and  
$|v_1\rangle=\frac{(1+i)}{2}\bigl(\begin{smallmatrix}
%1\\ i
-i \\ 1
\end{smallmatrix} \bigr)$. However, if the state vector $|v_0\rangle$ or $|v_1\rangle$ %(also referred to as intermediate signals in literature)
is applied to the control of a two-qubit gate, then the resulting output vector results in an entangled state~\cite{HSY+:2006}. % If a cascades of quantum primitives generates any output state as an entangled state, then the cascade can no longer be used to realize a binary reversible function. 
%Note that  direct synthesis methods of quantum circuits using quantum primitives must take into account this condition. However, 
If a quantum circuit is obtained from the quantum decomposition of a MCT circuit, the entangled state does not arise. % at any intermediate position and clearly the circuit realizes a binary function. 
%%%%%%%%%%%%%%%%%%%%%%%%%%%%%%%%%%%
% it was not in previous version
\begin{definition}\label{def:1}
 If a quantum circuit generates an entangled state for any given binary input state is said to be an entangled circuit. 
\end{definition}
\begin{example}
The cascades of quantum primitives shown in Figure~\ref{fig:engt_lnnC}
 is an entangled circuit because  the circuit generates an entangled state for input vector $\langle 1,1,1 \rangle$ and the resulting outputs are not separable into $3$ single-qubit states.
\end{example}

\begin{figure}
  \centering
 \input{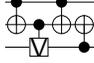}
\caption{An entangled circuit.}
\label{fig:engt_lnnC}
\end{figure}

%A quantum circuit with non-adjacent control and target of 2-qubit gates is referred to as a {\it standard quantum circuit}. However, if the control and target of 2-qubit quantum  gate are placed anywhere in a standard circuit model then the physical implementation of such type circuits is impractical due to the highly limited resources--qubits. 

%// modify this line: A quantum circuit that contains gates where the controls and targets lines are not necessaryly adjecent, is referred to as a {\it standard quantum circuit}.
A quantum circuit that contains gates which are not necessarily acting on the adjacent qubits, is referred to as a {\it standard quantum circuit}. A Linear Nearest Neighbor (LNN) quantum circuit is defined as follows:
%%%%%%%%%%%%%%%%%%%%%%%%%%%%%%%%%%%%%%%%%
\begin{definition}\label{def:2}
A quantum circuit $C$ is said to be a LNN circuit if all gates are acting on adjacent qubits.
\end{definition}

\begin{definition}\label{def:3}
The {\bf cost} of a circuit $C$ is defined as the number of its gates and denoted by $|C|$. For a given function $f$, a circuit $C$ is said to be optimal if there is no realization of $f$ with lower cost.
\end{definition}

The best reported LNN realization of the $T_3$ gate has quantum cost $9$. However, different LNN realizations of $T_3$ with cost $9$ are possible by not only replacing Controlled-$V$ (Controlled-$V^{\dagger}$) with Controlled-$V^{\dagger}$ (Controlled-$V$) but also by using the two different realizations of the SWAP gate as shown in Figure~\ref{fig:k}\subref{fig:SWAP} and \subref{fig:SWAP2}.
\noindent
\begin{figure}
  \centering
 \subfigure[]{
\input{g.tex}
\label{fig:g}
}
\subfigure[]{
\input{SWAP.tex}
\label{fig:SWAP}
}
\subfigure[]{
\input{SWAP2.tex}
\label{fig:SWAP2}
}
\subfigure[]{
 \input{t3.tex}
\label{fig:t32}
}
  \subfigure[]{
\small
\input{t3.tex}
\label{fig:t3}
}
\caption{\subref{fig:g} symbol of SWAP gate, \subref{fig:SWAP} and \subref{fig:SWAP2} quantum realization of SWAP gate, \subref{fig:t32} $T_3$ and \subref{fig:t3} LNN implementation of $T_3$.}
\label{fig:k}
\end{figure}
\noindent
The synthesis flow for the generation of LNN circuits %basically follows the steps
is done in 3 steps:  $i)$  decomposition of a MCT circuit into a quantum circuit, $ii)$ transformation of the resulting gates into LNN architecture by inserting SWAP gates or appropriate SWAP sequences and  $iii)$ optimization of the LNN circuits with post-synthesis methods~\cite{Robert:2011, Mper:2011, ZMismvl+:2011}. In this straightforward implementation, the resulting LNN circuits might be entangled realizations and suboptimal. For example, the circuit shown in Figure~\ref{fig:lnnTof31} \subref{fig:t31q} is an optimal standard quantum realization of the circuit shown in Figure~\ref{fig:lnnTof31} \subref{fig:t31}.     
By inserting SWAP gates to move the control of both CNOT towards the target results in a LNN circuit with quantum cost $17$. The insertion of appropriate SWAP sequences results in a circuit with quantum cost $13$ as shown in Figure~\ref{fig:lnnTof31} \subref{fig:t31qlnn}. However, the circuit is an entangled circuit 
%according to the definition~\ref{def:1} 
and we ignore such type of realization. Moreover, for the MCT circuit as shown in Figure~\ref{fig:opt_rbtexact} \subref{fig:r1}, the optimization method proposed in~\cite{Robert:2011} results in a LNN circuit with quantum cost $24$ (Figure~\ref{fig:opt_rbtexact} \subref{fig:rexact}). By replacing the SWAP gates with appropriate SWAP sequences as proposed in~\cite{ZMismvl+:2011} the circuit with cost $18$ as shown in Figure~\ref{fig:opt_rbtexact} \subref{fig:apswap} is obtained.
This circuit is not minimal. %, since it can be further optimized with our proposed LNN templates. 
 %How it is, can be found section~\ref{sec:optlnn}.  The LNN  realization of Toffoli-$3$ as shown in Figure~\ref{fig:k}\subref{fig:t3} is used to transform MCT circuits into LNN architecture in this literature. 
 
\begin{figure}
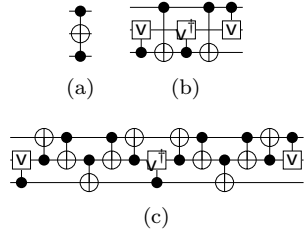

  \centering
\subfigure[]{
 \input{t31.tex}
\label{fig:t31}
}
  \subfigure[]{
\small
\input{t31q.tex}
\label{fig:t31q}
}

\subfigure[]{
\small
\input{engt32.tex}
\label{fig:t31qlnn}
}
\caption{\subref{fig:t31} $T_3$ with non-adjacent controls, \subref{fig:t31q} optimal quantum realization of \subref{fig:t31} and \subref{fig:t31qlnn} LNN implementation of~\ref{fig:t31q} with cost $13$.}
\label{fig:lnnTof31}
\end{figure}
 
\begin{figure}
  \centering
\subfigure[]{
 \input{robex.tex}
\label{fig:r1}
}
\subfigure[]{
\small
\input{robexexact.tex}
\label{fig:rexact}
}
\subfigure[]{
\small
\input{apswap.tex}
\label{fig:apswap}
}

\caption{\subref{fig:r1} A MCT circuit, \subref{fig:rexact} its LNN implementation according to~\cite{Robert:2011} and \subref{fig:apswap} and as proposed in~\cite{ZMismvl+:2011}.}
\label{fig:opt_rbtexact}
\end{figure}

% quantum gates and therefore, the size of LNN circuits obtained from transforming standard circuits by using SWAP gates is significantly high. However, we propose LNN implementations without using SWAP gate results in considerably reduced circuits can be discussed subsequently. 
%Nearest Neighbor Cost (NNC)--- If a 2-qubit quantum gate g where its control and target are placed at the $c^{th}$ and $t^{th}$ line $(0 \leq c; t < n)$, respectively. The NNC of g is defined as $|c -t -1|$ , i.e. distance between control and target lines. The NNC of a circuit is defined as the sum of the NNCs of its gates. Optimal NNC for a circuit is 0 where all quantum gates are either 1-qubit or 2-qubit gates performed on adjacent qubits[robert paper].
%In general, a \textbf{quantum template} is  defined as a quantum identity circuit that can not be reduced by other templates~\cite{MYDM:2005}. If a sequence of gates in circuits to be optimized matches a sequence of gates in a template, then it can be replaced with the inverse  of the remaining sequence of the template. This process is known as \textbf{template matching}. At the beginning, the basic quantum identities are considered as templates to be applied in the template matching process.  The \textbf{reconfigured templates} can be derived from theses templates by using splitting rules and the significance of reconfigured templates has been shown in~\cite{MDA+:2011}. 
\end{section}

\begin{section}{Transformation of MCT Circuits into LNN Circuits}\label{sec:Trans}
 In this section, we propose methods for transforming MCT circuits into LNN architecture by using three different models to move the control (target) of a $2$-qubit quantum 
 %(generalized Toffoli) 
gate towards the target (control) until they become adjacent. This approach always results in non-entangled LNN circuits with considerably lower quantum cost than previously proposed methods. 

\begin{subsection}{LNN Transformation of 2-qubit Quantum Gates}\label{cnot}
For a standard quantum circuit, the Nearest Neighbor Cost (NNC) of a 2-qubit quantum gate $g$, where its control and target are placed at the $c^{th}$ and $t^{th}$ line 
%$(0 \leq c; t < n)$ 
respectively, is defined as $|c -t| -1$, i.e. the distance between control and target lines~\cite{Robert:2011}. 
%It has been observed that 
The $CNOT$ gate with $NNC=1$ as shown in Figure~\ref{fig:trans}\subref{fig:cnot} has three different LNN implementations as shown in Figure~\ref{fig:trans}\subref{fig:lnnD}, \subref{fig:lnnF} and \subref{fig:lnnI} that  we  refer to as \textbf{Model-1}, \textbf{Model-2}, and \textbf{Model-3} respectively. Clearly, fewer gates are needed in each LNN implementation than with SWAP gates. This model can be generalized for $NNC=k$ as follows:
\begin{figure}
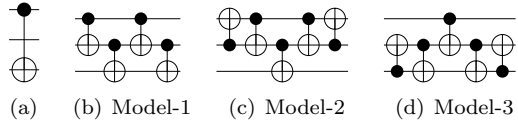

  \centering
\subfigure[]{
 \input{cnot.tex}
\label{fig:cnot}
}
  \subfigure[Model-1]{
\input{D.tex}
\label{fig:lnnD}
}
  \subfigure[Model-2]{
\input{F.tex}
\label{fig:lnnF}
}
  \subfigure[Model-3]{
\input{I.tex}
\label{fig:lnnI}
}
\caption{LNN transformation of $2$-qubit quantum gates with non-adjacent control and target}
\label{fig:trans}
\end{figure}

\textbf{Model-1 (Control moves towards target):} A $CNOT$ gate with $NNC=k$ in $n$-qubit circuit $1\leq k <n-1$ can be transformed into a LNN architecture with quantum cost $4k$ by using this model whereas it requires $6(k+1)$ quantum gates if SWAP gates are used. % in LNN transformation. 
For instance, the $CNOT$ with $NNC=4$ and its LNN transformation to move the control towards the target as shown in Figure~\ref{fig:t26aa}\subref{fig:t2611}, \subref{fig:t261a}, \subref{fig:t261b} and \subref{fig:t261d}. The $2^{nd}$ and $4^{th}$ $CNOT$ gates in Figure~\ref{fig:t26aa}\subref{fig:t261a} are replaced with their reverse implementation of each other by using Model-1. 
The resulting circuit is shown in Figure~\ref{fig:t26aa}\subref{fig:t261c}.
This process is iterated until no $CNOT$ gates with $NCC > 0$ remain. The final circuit is shown Figure~\ref{fig:t26aa}\subref{fig:t261d}.
%successively.    
\begin{figure}
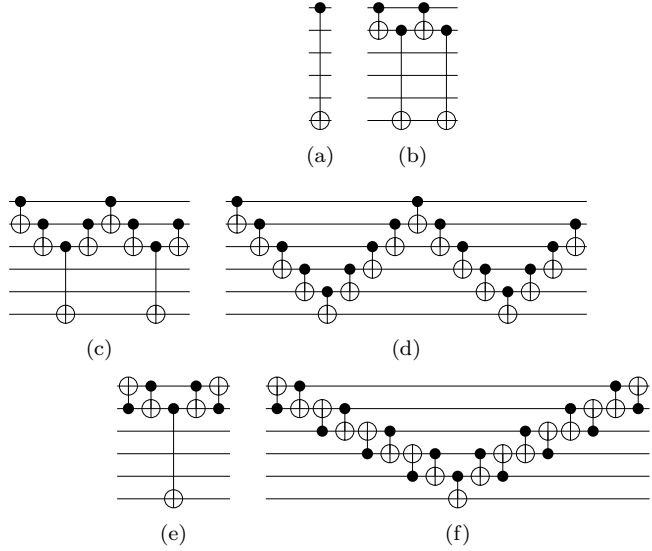

  \centering
\subfigure[]{
 \input{t261.tex}
\label{fig:t2611}
}
  \subfigure[]{
\input{t261a.tex}
\label{fig:t261a}
}

  \subfigure[]{
\input{t261b.tex}
\label{fig:t261b}
}
\subfigure[]{
\input{t261d.tex}
\label{fig:t261d}
}
$\newline$
  \subfigure[]{
\input{t262.tex}
\label{fig:t262}
}
\subfigure[]{
\input{t265.tex}
\label{fig:t265}
}
\caption{LNN transformation of $CNOT$ with NNC=4.}% (16 gates in Model-1 and 17 gates in Model-2).}% by using Model-1 (16 gates).}
\label{fig:t26aa}
\end{figure}

\textbf{Model-2 (Control moves towards target):} A $CNOT$ with $NNC=k$ in $n$-qubit circuit $1\leq k <n-1$ can be transformed into a LNN architecture with quantum cost $4(k+1)$ by using Model-2.  % into LNN architecture. 
For instance, the $CNOT$ with $NNC=4$ can be transformed to a LNN circuit by iteratively moving the control towards the target as shown in Figure~\ref{fig:t26aa}\subref{fig:t262} and~\subref{fig:t265}.
This model can also be used for transforming Controlled-$V$ and Controlled-$V^\dagger$ gates with non-adjacent control and target lines.
\noindent

\textbf{Model-3 (Target moves towards control):} This model can be used to move the target to the control of a $CNOT$ with $NNC=k$. This transformation requires $4(k+1)$ gates. 
\end{subsection}

In summary, Controlled-$V$ or Controlled-$V^{\dagger}$ with non-adjacent control and target can only be transformed by using Model-2. Model-2 and Model-3 can be used to move controls (target) towards the target (controls) of a MCT gate. Model-1 enables the move of the control towards the target in $CNOT$ gates.

\begin{subsection}{LNN Transformation of Toffoli Gates}
$T_3$ gates with non adjacent controls and target can be transformed into MCT circuits where all gates have adjacent controls. Two different cases can be considered.
 
 Let $p, q$ be the number of free lines in between the controls $C=\{c_1,c_2\}$ ($c_1 < c_2$) and the target $t$ of $T_3(C,t)$, then the following 2 cases are possible.
 
 \textbf{Case 1:}  If $c_1=i$, $c_2=i+q+1$ and $t=c_1-p-1$ or $t=c_2+p+1$ and $0\leq p, q$ 
%if two controls are adjacent but target is non-adjacent to controls, 
then the control $c_1$ can move towards the $c_2$ and the target $t$ can move towards the control $c_2$ by using $4(p+q)$ gates or the control $c_2$ can move towards the $c_1$ and the target $t$ can move towards the control $c_1$ by using $4(p+q)$ gates results in a LNN circuit with $4(p+q)+9$ gates. When $q=0$, the controls are adjacent, for instance the%Figure~\ref{fig:t3_4_a_b} and
~\ref{fig:t3666}\subref{fig:t3_6_c} shows the form of transformation $T_3$ with $6$ lines when $q=0$ and $p=3$. The replacement of $T_3$ with its LNN circuit results in a LNN architecture of~\ref{fig:t3666}\subref{fig:t3_6}.
\noindent
\begin{figure}
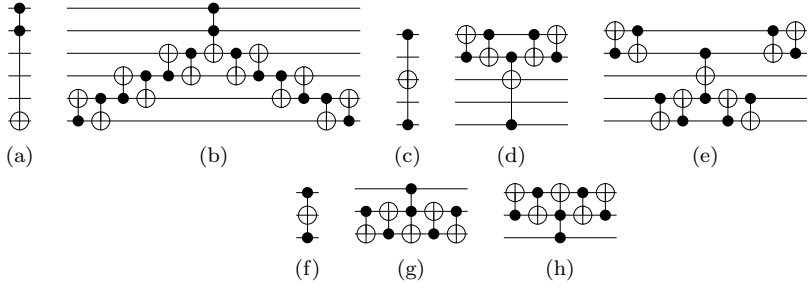

  \centering
\subfigure[]{
 \input{t3_6.tex}
\label{fig:t3_6}
}
\subfigure[]{
\input{t3_6_c.tex}
\label{fig:t3_6_c}
}
   \subfigure[]{
\input{t3_5.tex}
\label{fig:t3_5}
}
\subfigure[]{
\input{t3_5_a.tex}
\label{fig:t3_5_a}
}
\subfigure[]{
\input{t3_5_b.tex}
\label{fig:t3_5_b}
}
$\newline$
\subfigure[]{
 \input{t31.tex}
\label{fig:t323}
}
  \subfigure[]{
\input{t31_b.tex}
\label{fig:t31_b}
}
 \subfigure[]{
\input{t31_a.tex}
\label{fig:t31_a}
}
\caption{ LNN transformation of $T_3$.}
\label{fig:t3666}
\end{figure}
\noindent

\textbf{Case 2:} If $c_1=t-p-1$, $c_2=t+q+1$, $0\leq p, q$ then the the controls can move towards the target by using $4(p+q)$ gates.
When $p=q=1$, $T_3$ is the form as shown in Figure~\ref{fig:t3666}\subref{fig:t3_5}. Two controls can move towards the target as shown in Figure~\ref{fig:t3666}\subref{fig:t3_5_a} and   \subref{fig:t3_5_b} successively. When $p=0$ and $q=0$ the $T_3$ as the form shown in Figure~\ref{fig:t3666}\subref{fig:t323}. Further, the two the controls can be adjacent as the form shown in Figure~\ref{fig:t3666}\subref{fig:t31_b} or \subref{fig:t31_a} by using 4 gates.  Therefore, the final LNN circuit requires $4(p+q+1)+9$ gates when $0<p,q$. By replacing $T_3$ in circuits~\ref{fig:t3666}\subref{fig:t31_b} and \subref{fig:t31_a} with its LNN implementation results in LNN architectures with $13$ gates. Moreover, the resulting LNN circuit of~\ref{fig:t3666}\subref{fig:t323} would be non-entangled whereas the previously published approach of LNN transformation gives entangled circuit in this case. However, if the $T_3$ in MCT circuits is either one of the form $T_3(c_1, c_2, t)$ or $T_3(t, c_1, c_2)$ before quantum decomposition of circuits then the synthesis flow of LNN circuits ensures the non-entangled LNN circuit as a result.   

\end{subsection}

\end{section}

\begin{section}{Optimization of LNN Circuits with LNN Templates}\label{sec:optlnn}
LNN circuits obtained from the proposed transformation of MCT circuits are most likely not minimal even if an optimal standard quantum circuit is transformed into an LNN architecture. For instance, by using the models proposed in Section~\ref{cnot}, the three different LNN implementations shown in Figure~\ref{fig:std_lnnC}\subref{fig:std_lnnC1}, \subref{fig:std_lnnC2}, and \subref{fig:std_lnnC3} of the optimal standard quantum circuit shown in Figure~\ref{fig:std_lnnC}\subref{fig:s}. However, none of these implementations are minimal. 

\begin{figure}
  \centering
\subfigure[]{
 \input{stdopt.tex}
\label{fig:s}
}
  \subfigure[]{
\input{std_lnnC1}
\label{fig:std_lnnC1}
}
\subfigure[]{
\input{std_lnnC2}
\label{fig:std_lnnC2}
}
\subfigure[]{
\input{std_lnnC3}
\label{fig:std_lnnC3}
}
\subfigure[]{
\input{opt}
\label{fig:std_opt}
}
\caption{LNN transformations of~\subref{fig:s}: \subref{fig:std_lnnC1} using model-1, \subref{fig:std_lnnC2} using model-2, \subref{fig:std_lnnC3} using model-3, and \subref{fig:std_lnnC1} optimized circuit.}
\label{fig:std_lnnC}
\end{figure}

The idea of post synthesis optimization -- template matching -- for simplifying standard MCT %(quantum) 
circuits originated in~\cite{MMD:2003} and later on extensive studies have been done by introducing reconfigured templates~\cite{MDA+:2011}, developing an algorithm to find templates~\cite{MDWCCI:2012} as well as modifying the definition of template and analizing their properties~\cite{MD+:2012}. Template matching  has been extended to optimize LNN circuits based on templates that are comprised of SWAP gates~\cite{Robert:2011}. % and still this approaches do not results minimal LNN circuits, why it is  explained in subsequent examples. 
In this section, we propose LNN templates that can be used in template matching to optimize LNN circuits. This new approach outperforms the previously proposed approaches.
% Template matching reduces the size of a circuit if a template is applied successfully. %regardless the control and target of a gate are adjacent in standard circuits, however, 
%
With LNN it is necessary to have quantum templates that ensure the resulting optimized circuit does not violate the constraint of LNN quantum circuits when a template is applied.  Therefore, we first present the formal definition of LNN templates.  The properties of templates proposed in~\cite{MD+:2012} hold for theses templates as well. 

\begin{definition}
A {\bf LNN quantum template} is an LNN identity circuit with $d$ gates, such that at least one sequence of ${\lfloor \frac{d}{2} \rfloor+1}$ gates in the circuit  can not be reduced by any other LNN template.
\end{definition}

Clearly, all two-qubit templates as well as all templates proposed in~\cite{MD+:2012} for which the LNN constrain holds, must be the LNN templates. The significance of proposed LNN templates shown in Figure~\ref{fig:lnnT3q} is illustrated with the subsequent examples.    
\noindent 
\begin{figure}
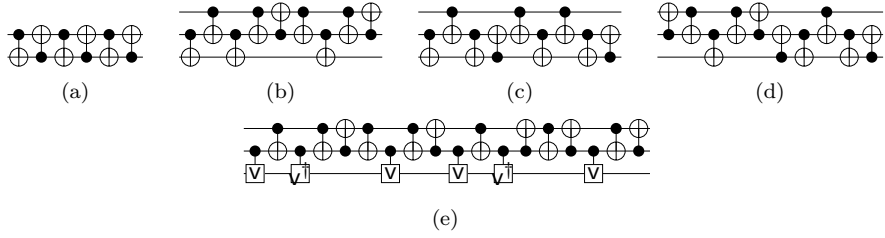

  \centering
  \subfigure[]{             
 \input{tm4}
\label{fig:2tm6}
}
  \subfigure[]{
\input{lnnT_gdf.tex}
\label{fig:lnnT_gdf}
}
  \subfigure[]{
\input{lnnT_gdi_2.tex}
\label{fig:lnnT_gdi}
}
\subfigure[]{
\input{lnnT_fdgi.tex}
\label{fig:lnnT_fdgi}
}
\subfigure[]{
\input{t_18.tex}
\label{fig:t_18}
}
\caption{LNN quantum templates.}
\label{fig:lnnT3q}
\end{figure}
\noindent

\begin{example}\label{ex:1}
The gate sequence in the LNN circuits shown in Figure~\ref{fig:std_lnnC}\subref{fig:std_lnnC1} and \subref{fig:std_lnnC2} match with the templates in Figure~\ref{fig:lnnT3q}\subref{fig:lnnT_fdgi} and \subref{fig:lnnT_gdf}. Template matching results in an optimized circuit as shown in Figure~\ref{fig:std_lnnC}\subref{fig:std_opt}.
 These small circuits cannot be optimized by previously proposed methods.  
\end{example}

\begin{figure}
  \centering
\subfigure[]{
 \input{robex.tex}
\label{fig:r}
}
\subfigure[]{
\input{robex7}
\label{fig:r7}
}
\subfigure[]{
\input{apswapopt}
\label{fig:apswapopt}
}
\caption{\subref{fig:r} MCT circuit, \subref{fig:r7}Optimized LNN circuit of~\subref{fig:r} and \subref{fig:apswapopt} Optimized circuit in Figure~\ref{fig:opt_rbtexact}\subref{fig:apswap}. }
\label{fig:opt_rbt}
\end{figure}

\begin{example}\label{ex:2}
Consider the circuit in Figure~\ref{fig:opt_rbt}\subref{fig:r} reported in~\cite{Robert:2011}. According to our proposed approach, %the LNN transformation and optimization have been shown in Figure from~\ref{fig:opt_rbt}\subref{fig:r3} to \ref{fig:r7}.
the LNN transformation and optimization %of circuit in Figure~\ref{fig:opt_rbt}\subref{fig:r}  
are done by the steps: 1) move targets towards the controls by using Model-3, 2) replace $T_3$ with its LNN circuit, 3) apply gate deletion rules, 4) apply template \ref{fig:lnnT3q}\subref{fig:lnnT_fdgi}, and 5) apply gate merge rules~\cite{MDWCCI:2012}. The resulting optimized circuit is shown in Figure~\ref{fig:opt_rbt}\subref{fig:r7}.
The number of quantum gates in the optimized circuit is 13. % whereas the simplified circuit with quantum cost $24$ is shown in Figure~\ref{fig:opt_rbtexact}\subref{fig:rexact}. 
The cost of the solution proposed in~\cite{Robert:2011}  is almost  50\% higher (see Figure~\ref{fig:opt_rbtexact}\subref{fig:rexact}).
However, the proposed templates in~\cite{Robert:2011} are derived from SWAP gates, therefore, the resulting LNN circuit is still contains SWAP gates. % with no-minimal realizations that can be further optimized. Moreover, 
The optimization by choosing  appropriate SWAP sequence proposed in~\cite{ZMismvl+:2011} results a circuit with cost $18$ as shown in Figure~\ref{fig:opt_rbtexact}\subref{fig:apswap}. However, the gate sequence from index $4$ (starting at $0$) to $14$ and further reconfiguring $16^{th}$ of the template as shown in Figure~\ref{fig:lnnT3q}\subref{fig:t_18}  matches with the gate sequence $0,1,4,5,6,7,8,9,10,11,12,13$ in circuit in Figure~\ref{fig:opt_rbtexact}\subref{fig:apswap}. Therefore, template matching results in a circuit with cost $13$ as shown in Figure~\ref{fig:opt_rbt}\subref{fig:apswapopt}. %It is clear that the later approach takes %quite longer 
\end{example}
\end{section}

\begin{section}{3-Qubit Optimal LNN Circuits}\label{sec:lnnopt}
%Considering the technological limitations, synthesis of optimal LNN circuits is a must. 
In general, the direct synthesis of quantum circuits for a given reversible function specification is intractable. However, for $3$-qubit functions, all optimal standard quantum circuits have been obtained by directly cascading the quantum primitives~\cite{ISM+:2012}. Therefore, a similar method can be used to find all optimal LNN circuits of 3 qubits. 

\begin{definition}
Given a library of gates $L$, a LNN circuit $c$ with $n$ gates that realizes the function $f$, is said to be optimal with respect to $L$, if no LNN realization of $f$ exists that has fewer than $n$ gates.
\end{definition}

Let $C_n$ be the set of all optimal circuits with $n$ gates.
In constructing LNN circuits, we use the $15$ permuted quantum gates with 3 qubits whose control and target are acting on the adjacent qubits. An exhaustive search method has been used to find all LNN quantum circuits $C_n$ by cascading the optimal LNN quantum circuits from the sets $C_{n-1}$  and $C_1$. 
For all 3-qubit binary functions, the results of optimal LNN quantum circuits are shown in column II  in Table~\ref{tab:TOPT}.
\end{section}

\section{Synthesis Flow of LNN circuit}
%Direct synthesis of quantum circuits are intractable, even it is very hard to achieve an optimal standard quantum circuit of higher-order Toffoli gate. The 
LNN decomposition of Higher-Order Tofolli gates has been studied in~\cite{ZMismvl+:2011} in which the minimized standard quantum circuit of Higher-Order Toffoli is transformed into a LNN circuit by inserting appropriate SWAP gates. However,  it is evident that the insertion of  SWAP gates into optimal standard quantum circuit results LNN circuits that can still be optimized. We investigate  the minimal way of transforming Higher-Order Toffoli gate into LNN architecture in which optimization is to be done at the end of the process. The synthesis flow of LNN circuit is shown in Algorithm~\ref{trans:algo1}.   
\begin{algorithm}
\caption{Synthesis flow  LNN circuit}
\label{trans:algo1}
\begin{algorithmic}
\STATE 1) Decompose Higher-Order Toffoli in a MCT circuit into $T_3$ gates according to~\cite{BBC+:1995}.
\STATE 2) Transform all $T_3$ gates with non-adjacent controls and target by using Model-2 and Model-3 results in a circuit of all Tofolli-$3$ with adjacent controls and target.
\STATE 3) Replace all $T_3$ with its LNN architecture results in a non-minimal LNN circuit.
\STATE 4) Optimize the circuit obtained in step 3 by using LNN quantum templates.    
\end{algorithmic}
\end{algorithm}

\begin{figure}
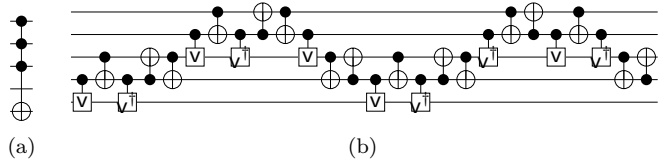

  \centering
   \subfigure[]{
\input{t4.tex}
\label{fig:t4}
}
\subfigure[]{
\small
\input{t4_15.tex}
\label{fig:t4_15}
}
%\caption{Optimization of LNN circuit for High-Order Toffoli}
\caption{ Optimized LNN circuit of $T_4$ with one extra line.}
\label{fig:t3555}
\end{figure}

According to~\cite{BBC+:1995}, to transform a Higher-Order Toffoli gate into a circuit with $T_3$ requires at least one extra line, 
however, the decomposition by using more lines results in a circuit with the less number of $T_3$. We observed that the above synthesis flow gives better results if more working lines are used in decomposition. We achieve the LNN circuit as shown in Figure~\ref{fig:t3555}\subref{fig:t4_15} for $T_4$ with one working line.
 
\section{Experimental Results}\label{section:exp}

\begin{table}
  \centering
\tiny
\begin{threeparttable}  
    \begin{tabular}{|r|r|r|r|r|r|r||r|r|r|r|r|r|r|}\hline
    Size     & LNN & S~\cite{Robert:2011}    & AS~\cite{ZMismvl+:2011}  & MS  & M     &  Opt(M)     & Size    & LNN & S~\cite{Robert:2011}     & AS~\cite{ZMismvl+:2011}  & MS  & M     & Opt(M) \\ \hline
    0          & 1     & 1     & 1     & 1     & 1     & 1      & 32    &              & 786   & 198   & 1365  & 1461  &1131  \\\hline
    1         & 7     & 7     & 7     & 7     & 7     &  7     & 33    &              & 1442  & 1076  & 566   & 1728  &825  \\\hline
    2        & 29    & 29    & 29    & 29    & 29    &29    & 34    &              & 2899  & 3267  & 1701  & 2056    &577  \\\hline
    3        & 82    & 74    & 74    & 74    & 74    &80       & 35    &              & 2212  & 2409  & 1515  & 1491  &463  \\\hline
    4       & 181   & 110   & 110   & 112   & 112   &169       & 36    &              & 1879  & 176   & 1790  & 649   &340  \\\hline
    5       & 334   & 100   & 102   & 120   & 120   &307       & 37    &              & 1587  & 392   & 1616  & 1631  &176  \\\hline
    6      & 374   & 33    & 53    & 125   & 125   &324       & 38    &              & 276   & 1910  & 625   & 1870  &100  \\
\hline
    \end{tabular}%
%\begin{tablenotes}
\small
{\bf Size:} Number of gates in a circuit,
{\bf LNN:} Number of minimal $3$-qubit LNN circuits,  
{\bf S:} Transformation of minimal MCT circuits by using SWAP gates proposed in~\cite{Robert:2011},  
{\bf AS:} Transformation of minimal MCT circuits by using appropriate SWAP sequence  proposed in~\cite{ZMismvl+:2011},  
{\bf MS:} Transformation of minimal MCT circuits by using model-1 and SWAP gates,
{\bf M:} Transformation of minimal MCT circuits by using proposed models and 
{\bf Opt(M):} Optimized results obtained from LNN circuits in coulmn $M$.
\caption{Transformation and optimization of LNN circuits.}  
\label{tab:TOPT}%
\end{threeparttable}
\end{table}%
The proposed synthesis flow of LNN circuits has  been implemented in C/C++ on top of RevKit-1.2.1 \cite{RevKit}. % tools for reversible circuit design. 
To evaluate the effectiveness of the new approach, we have taken all 3-qubit minimal MCT circuits and transformed them into LNN circuits by using different approaches as shown in columns III, IV,  V, and VI of Table~\ref{tab:TOPT}.  The proposed transformation approach results the average number of gates 27.1 %as shown in column M whereas it is 15.9 in optimal LNN circuits. 
compared to the optimal of 15.9.
It can be seen that the new transformation method results in smaller circuits. The results of optimized LNN circuits in column Optz(M) are obtained by template matching using 21 LNN templates. The results shows that $41\%$ gate reduction is required on average to reach the optimal LNN circuits shown in column II. However, we gain an approximate $19\%$ reduction and a further $27\%$ reduction is needed for the optimal result.

\section{Conclusion} \label{section:Conclusion}
%In comparing with the optimal standard quantum circuits, the optimal LNN circuits are costly but reasonable to resolve the limitations in quantum technology. 
We propose a new synthesis flow for LNN quantum circuits in which the transformation models result in circuits with considerable lower quantum cost compared to others methods. Moreover, the template matching with new LNN templates significantly reduces the number of gates in circuits. In some cases, the reduction is more than $50\%$. 
The effectiveness of our approach is evident in the examples.

\bibliographystyle{IEEEtran}
\bibliography{lit_myrev}
\end{document}